# Mildly-Doped Polythiophene with Triflates for Molecular Recognition

Aicha Boujnah[a], Aimen Boubaker[a], Adel Kalboussi[a], Kamal Lmimouni[b] & Sébastien Pecqueur[b]*

[a] Department of Physics, University of Monastir Tunisia.

[b] Univ. Lille, CNRS, Centrale Lille, Univ. Polytechnique Hauts-de-France, UMR 8520 - IEMN, F-59000 Lille, France.

Email: sebastien.pecqueur@*iemn*.fr

## Abstract

**Organic semiconductors have enough molecular versatility to feature chemo-specific electrical sensitivity to large families of chemical substituents via different intermolecular bonding modes. This study demonstrates that one single conducting polymer can be tuned to either discriminate water-, ethanol- or acetone-vapors, on demand, by changing the nature of its dopant. Seven triflate salts differ from mild to strong p-dopant on poly(3-hexylthiophene) sensing micro-arrays. Each material shows a pattern of conductance modulation for the polymer which is reversible, reproducible, and distinctive of other gas exposures. Based on principal component analysis, an array doped with only two different triflates can be trained to reliably discriminate gases, which re-motivates using conducting polymers as a class of materials for integrated electronic noses. More importantly, this method points out the existence of tripartite donor-acceptor charge-transfer complexes responsible for chemospecific molecular sensing. By showing that molecular acceptors can have duality to p-dope semiconductors and to coordinate donor gases, such behavior can be used to understand the role of frontier orbital overlapping in organic semiconductors, the formation of charge-transfer complexes via Lewis acid-base adducts in molecular semiconductors.**

## Introduction

Sensors and molecular recognition are two different paradigms in which the boundaries of the one cannot contain the definition of the other: Molecules do not belong to a continuum, but a space-group of discrete-objects that reaches already ten million synthetically-plausible entities with less than 17 atoms and nine elements of the periodic table (FDB-17).[**Ruddigkeit2012**][**Bühlmann2020**] Therefore, finding many sensors that selectively quantifies each molecule among all others is far less trivial than quantifying a physical property that obeys a known deterministic model. To overcome this conflict entangling molecular sensing's technology-readiness-level, statistical approaches are adopted to directly infer on real-world environments by mean of learning/discrimination rather than calibrating/ quantification. In this scope, lots of electronic nose technologies have been identified [**Persaud2002**][**Röck2008**][**Wilson2009**][**Karakaya2020**][**Kim2021**], that have been employed in various fields: medicine [**Turner2004**][**Farraia2019**] [**Wulandari2020**], military [**Nagappan2017**], agronomy, [**Tan2020**] food-quality [**Loutfi2015**][**Falasconi2012**], pharmaceutical [**Wasi-lewski2019**] and environmental monitoring [**Wilson-2012**]. By analogy with the human olfactory system, such machine-learning based instruments combine sensitive array materials and pattern recognition techniques to classify odors as rich molecular imprint on the system inputs.[**Hu2019**] While sensors' figure-of-merits aim ranking hardware technologies for the widest application range,[**IEEE**] sensing-array pattern recognition's are intimately bound to their algorithm involved for specific classification tasks.[**Peveler2016**] [**Hierlemann2008**] Regarding hardware, various transducers can be used: electrical, optical (as photo-spectrometry [**Li2016**][**Li2018**][**Guo2020**], or surface-plasmon resonance - SPR),[**McDonagh2008**][**Maho-2020**][**Gaggiotti2020**] or mechanical (such as surface-acoustic-wave – SAW and quartz-crystal-microbalance - QCM).[**Chang2000**][**Vashist2011**][**Wilson2012**] Electrical transduction offers greater advantages for system integration with complementary metal-oxide semiconductor (CMOS) analog-front-end systems enabling amperometric, potentiometric, conductimetric or impedimetric sensing. [**Simic-2013**][**Ragogna2021**][**Ouh2018**] In such case, bipolar or field-effect transistor devices can integrate various kind of electrically conductive materials [**Swager1994**][**Park2019**]: [**Lv2017**] metal-organic-frameworks,[**Li2020**][**Chidambaram2018**] nanoparticles,[**Liu2019**][**Cattabiani2018**][**Peng2009**] nanotubes,[**Pandhi2020**][**Schroeder2018**][**Shulaker-2017**] two-dimensional semiconductors, [**Pandhi2020**][**Zhang2019**][**Mackin2018**][**Huang2021**] [**Kim2017**] or (macro)molecules [**McQuade2000**] [**Holliday2005**][**Janata2003**][**Lange2008**]. It is admitted that conducting polymers are the class that can theoretically show the highest sensitivity thanks to analyte permeation at the level of the molecular chains, to chemospecifically interact with the bulk accordingly to the molecular nature of the conducting material.[**Wang2020**][**Cicoira2008**][**Berggren2019**] [**Namsheer2021**] The ability for polar gases to alter the mobility due to swelling of polythiophene, polypyrole or polyaniline derivatives has been proposed as a physical mechanism driving the reversible conductance change.[**Tiggemann2017**][**Chang2006**][**Persaud2005**] [**Lizarraga2004**] Recent trends to increase their selectivity is conducted by mean of molecular imprinting a target gas upon electrodeposition of the polymer on the device [**Sharma2012**][**Cieplak-2016**][**Sharma2012**], creating molecular voids in the material using the steric hindrance of any molecules other than the molecular template. Various strategies in the synthetic-design of target-specific conducting polymers can enable high chemoselectivity [**Lu2020**][**Weis2015**][**Song2010**]. However, the iterative orthogonal co-integration for each soft chemosensitive polymers is inherently limiting additive-manufacturing yields for highly-dense multi-sensor micro-arrays [**Hoefflinger2020**].

As molecular donor/acceptor materials, any conducting polymer forms adducts with other molecules, to form a charge-transfer complex with frontier orbitals energy shallow with the semiconductor's density of states.[**Salzmann2012**] [**Salzmann2016**][**Méndez2013**][**Méndez2015**]. Their presence has a direct impact on the semiconductor doping and its conductivity, chemo-specifically with the dopants' molecular structure. For p-doping, molecular acceptors can have a Lewis acidic nature for which their electrophilicity can be rich enough to interact with the semiconductor environment [**Schmid2014**][**Pecqueur2016**][**Kellermann2015**], and also with volatile organic compounds for with the chemospecificity relate to the ligands chemistry between the p-dopant and the electron donors [**Cotton2005**][**Kugel2004**][**Reiland2000**].

In this study, this unique feature for conducting polymers is evidenced to enable molecular recognition by organic semiconductor doping. Chemospecificity of a polymer's conductance by the nature of its p-dopant is assessed to interact differently with water, acetone or ethanol. The support of principal component analysis (PCA), demonstrates their applicability for designing the selectivity of polythiophene at will by just changing the nature of the triflate dopants used in an array. PCA is a mathematical transformation that reduces data's dimensionality while preserving a maximum of its variance.[**Bishop2006**][**Bedoui2018**] By removing correlated components in the material's sensing response, this method allows finely defining parametric rules for the sensing materials calibration, to allow good classification of the sensed gas patterns by an algorithm. Reciprocally, PCA helped identifying the chemospecificity of each dopant for each gas, and highlights the existence of tripartite gas/dopant/semiconductor charge-transfer mechanisms ruling the materials conductivity.

## Results & Discussion

**Assessment of P3HT doping by triflate salts.**

A systematic comparison of different materials requires a unique deposition techniques for all elements, to assess property enhancements of the material and not to the process itself. Generic deposition processes are also a big concern for further technology readiness level increment for device manufacturing. However, the chemistry of triflate salts (small fluorinated/metal charged ions) and intrinsic polythiophenes (carbon-based neutral macromolecules) are by essence very different, which hinders their co-solubility in a unique solvent blend. This requires strong solvent compatibility (boiling point, volatility, dewetting) and adequately chosen deposition parameters (drying rate, time, temperature) to maintain the homogeneity during the process. For instance, we reported in the case of $Cu(OTf)_2$ homogeneously cosolubilized with the polymer, phase segregation between both materials can appear in the thin-film, leading to different grain boundaries in the materials' bulk.[**Ferchichi2020**] Here to compare the effects of the salts on the polymer, we choose to drop-cast the triflates on poly(3-hexylthiophene) (P3HT) top-contact devices by two sequential depositions instead of depositing both as one single blend, to better control the interfacing of the polymer with the triflates (Figure 1a-e).[**Jacobs2016**] As the process is spatially very heterogeneous, we studied the materials on small area (0.04 mm²) with micrometer-scaled conductimetric top-contact devices (616 µm² active area) clusterized as 16 conductimetric cells. The cells geometry have been design for the highest sensitivity with large width-to-length ratio (W/L) between $10^2$ (L = 2 µm) and $10^3$ (L = 400 nm) on 28 µm diameter devices, using interlaced spiral electrodes.[**Pecqueur2018b**]

Seven commercially available triflate salts have been tested on P3HT, for which we observed electrical performance enhancements that strongly depend on the salt, on log scale with the output current I (as previously reported with co-evaporated arylamine-based molecular semiconductors).[**Schmid2012**] In each case, such enhancement is observed and systematically compared before and after depositing the triflates on a same P3HT devices (Figure 1f). The nature of the current is clearly ohmic as $I \propto V^n$, with n = 1 and an applied voltage V ≤ 100 mV before and after triflate deposition. Because of the ohmic behaviour, the dopants' contribution in the current output can be used for low-voltage conductimetric sensing cell (output linear with the voltage), that requires much simpler analog-front-end readout circuits than amperimetric sensors (diode, transistors) in integrated consumer-electronics. Transmission line method shows that triflates enhance the conductance of the device by both a diminishing of the material's sheet resistance ($R_s$) and the device's contact resistance ($r_c$) (Figure 1f). Both enhancements are specific to doping, and testifies that triflates do not promote only the hole mobility of the polymer exclusively, but its density of charge carriers.[**Arkhipov2005**] It also shows that triflates promote both the conduction and the injection, suggesting that salts have diffused into the polymer (we experimentally observed that the annealing after deposition greatly helps for the device performances). For each material/device statistics, we generally observed a one-order-of-magnitude variability in the device resistance (R) characterized under nitrogen (Figure 1h). IV electrical characterization of the devices are very stable up to 100 mV (Figure 1h: 10 traces and retrace for each of the two acquisition series at 1 mV/s, error bars in Figure 1g). We evidence that most of the spreads are device dependent since deviations to the mean value generally remain identical before and after doping (as it can be seen in Figure 1gwhen comparing the relative position of the resistance measurements to the linear fitting). This indicates that most of the variability is essentially due to the intrinsic P3HT performance distribution and its deposition process (we see in Figure 1h that resistances are 1 to 2 fold distributed before doping), and not due to the dopant deposition. Despite the variability, figure 1h shows the trend of doping performance with the nature of the triflates, and particularly distinguishes stronger dopants (R < 10 kΩ in $N_{2(g)}$, ΔR/R between $10^{-3}$ and $10^{-5}$ with doping) from milder ones (R > 100 kΩ in $N_{2(g)}$, ΔR/R between 1 and 10% with doping). While the effect of $Cu(OTf)_2$ on P3HT has been reported,[**Ferchichi2020**] the one of $Bi(OTf)_3$ is novel for thiophene-based devices,[**Wemken2015**] with rather promising performances to p-dope photovoltaic cells',[**Mehmood2016**] or p-type transistors' contacts.[**Takamaru2021**] $Fe(OTf)_3$ was already known to be a source of triflate anions and efficiently undergo redox doping with poly(3,4-ethylenedioxythiophene).[**Gueye2016**] For the four other salts, the improvements in the device resistances are comparably moderate but substantial: At least 90% of the total charge carrier present in the P3HT is induced by the presence of the acceptors, and at least one order of magnitude improvements are observed. Although weak for organic electronic device efficiency, such mild doping remains completely relevant for sensing if used as a chemo-detector/electro-transducer in conductimetric sensing cells. The ordering of the dopant series by strength follows the one of redox potentials for their free cations in water, which will be discussed furthermore conjunctly with their chemo-sensitivity to molecular gases.

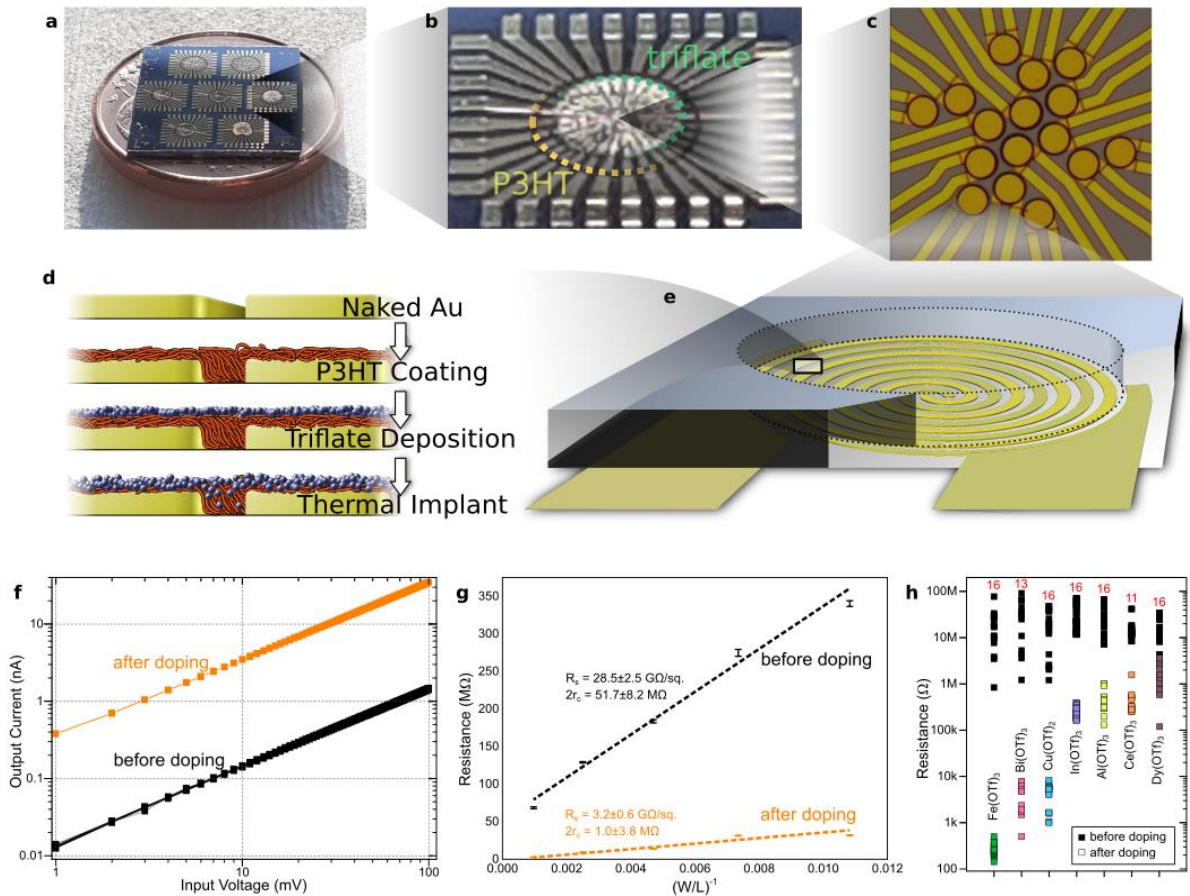

**Figure 1 | Device processing & doping material assessment. a,** Photograph of an e-beam lithographically-pattern electrode substrate, composed of seven arrays 16 conductimetric elements, clustered for the local deposition of sensitive materials by drop-casting. **b,** Zoom on one cluster of sensing elements, where the P3HT polymer and the triflate dopants are succesively depositied manually (here in this picture, over 5 and 4 mm$^2$). **c,** Optical microscope picture of a cluster of naked electrodes before depositing the sensitive materials (pitch: 30 μm). **d,** Sequencing for the deposition process to sensitize individually the clusters of 16 conductimetric elements with different triflate-doped P3HT bilayered materials. **e,** Schematic of the interdigitated concentric electrodes, for which the active area in contact with the sensitive material is specifically defined by the 28 μm diameter round aperture in the 2 μm thick passivation layer. **f,** Typical current-voltage electrical characteristic of a device measured before and after triflate drop-casting (example for Ce(OTf)$_3$). **g,** W/L dependency in the device resistance characterized before and after triflate drop-casting (example for Ce(OTf)$_3$). **h,** Resistance change observed for seven populations of devices, measured before and after the drop-casting of different triflates (evaluated on L = 400 nm devices).

.

**Sensitivity of triflated P3HT devices upon volatile gas exposure.**

Despite their stability under $N_{2(g)}$, we noticed many sources of conductance variations when measured or stored in air over time: First of all, the fact that materials are processed in air prior evaluation under $N_{2(g)}$ (with a 2-min curing at 70°C), testifies that the material does not irreversibly decompose with oxygen or moisture, showing redox stability of the triflate-doped P3HT coatings. When measured in air, an abrupt conductance change was observed with a gas flow (air or $N_{2(g)}$ with no distinctive difference between both), implying a first sensitivity degree of these materials to pneumatic stresses. All material have therefore been tested under static conditions at 1 mL/s (Figure 2a-c). Particularly for the three strongest dopants, namely $Fe(OTf)_3$, $Bi(OTf)_3$ and $Cu(OTf)_2$, air blowing on triflated P3HT tends to increase the current compared to prior turning-on the air flow. A possible explanation may originate from the hygroscopy for these strongest acceptors, for which the blow may displace the charge-transfer equilibrium between dopant and polymer by flushing the coordination sphere of the electron acceptor core, affecting the dopant's electrophilicity. Overall, the series of dopants by strength on P3HT evaluated in $N_{2(g)}$ (Figure 1g) remains comparably identical in air (Figure 2c). When the 1 mL/s flows through solvent-filled glass-vials, reversible changes in the P3HT materials are observed (Figure 2d-g). Tested for water, ethanol and acetone vapors, each gas imprints a specific pattern of positive and negative resistance modulation, which fully regenerated its pristine conductance state within the 3-min duration sequences that follows each gas exposure by an air purge. The observed modulations are specific from the gas and the triflated system, which promote either resistance increases or decreases. Figure 2d-f displays as an example the transient effect of the gases on a P3HT cell without triflate, with $Dy(OTf)_3$ (mild dopant) and with $Fe(OTf)_3$ (strong dopant) that clearly shows different dynamical patterns. Pristine P3HT does not show any sensitivity to ethanol, but +10% resistance modulations when exposed to water, and -15% resistance modulations when exposed to acetone. This shows that the P3HT itself has a moderate but genuine chemo-specific sensitivity to some gas (Figure 2d), as reported in the literature,[**Manoli2014**] and particularly that the response of polythiophene does not saturate with water at the opposite of polypyrole or polyaniline-based sensors.[**Matsuguchi2003**] [**Okuzaki2013**] When $Dy(OTf)_3$ is coated on P3HT, the polymer becomes sensitive to ethanol, and more interestingly the sensitivity toward acetone changes in an opposite way, as it increases the resistance and not decreases (Figure 2e). This shows that current modulations of the semiconductor are not just gas-specific, but interacts with the polymer through the participation of the dopant. For instance, molecular gas dipole moment has been proposed to be the indicator of a systematic diminishing of a polymer's resistance because of the material swelling which diminishes the charge carriers' mobility. Further, the hypothesis that gases directly dope the polymer through a redox process does not support that a same gas can either show resistance increase or decrease on the same polymer semiconductor, but doped by specific electron acceptors. In the case of $Fe(OTf)_3$, all gases induce a strong and distinctive effect on P3HT that is dedoped upon any gas exposure by 30-40% (Figure 2f). The change in conductance modulation signature with the dopant confirms the versatility to tune the gas sensitivity of the same polymer by tuning the dopant. More importantly, it shows that the whole salt participates in the charge-transfer rate with the polymer and that these salts do not systematically act as a source of triflate anions indifferently from the nature of the salt itself. Each of the three gases has a specific imprint on the different doped P3HT that is qualitatively reproducible for each of the six exposures per gases (Figure 2h), where it was observed that the trends for a given gas to dope/dedope P3HT is specific from the nature of the salt. The study performed for three devices for each of the eight materials was systematically made and the resistance modulation at the end of the 3min sequences was statistically compared for all the 24 devices in Figure 2i. Variability can be observed in the resistance modulations for devices with the same dopants, presumably due to the drop-casting deposition technique. However, qualitative trends of chemo-specific imprints are reproductively evidenced overall, particularly in the case of $Fe(OTf)_3$, $Cu(OTf)_2$, $Al(OTf)_3$, $Ce(OTf)_3$ and untriflated P3HT (Figure 2i).

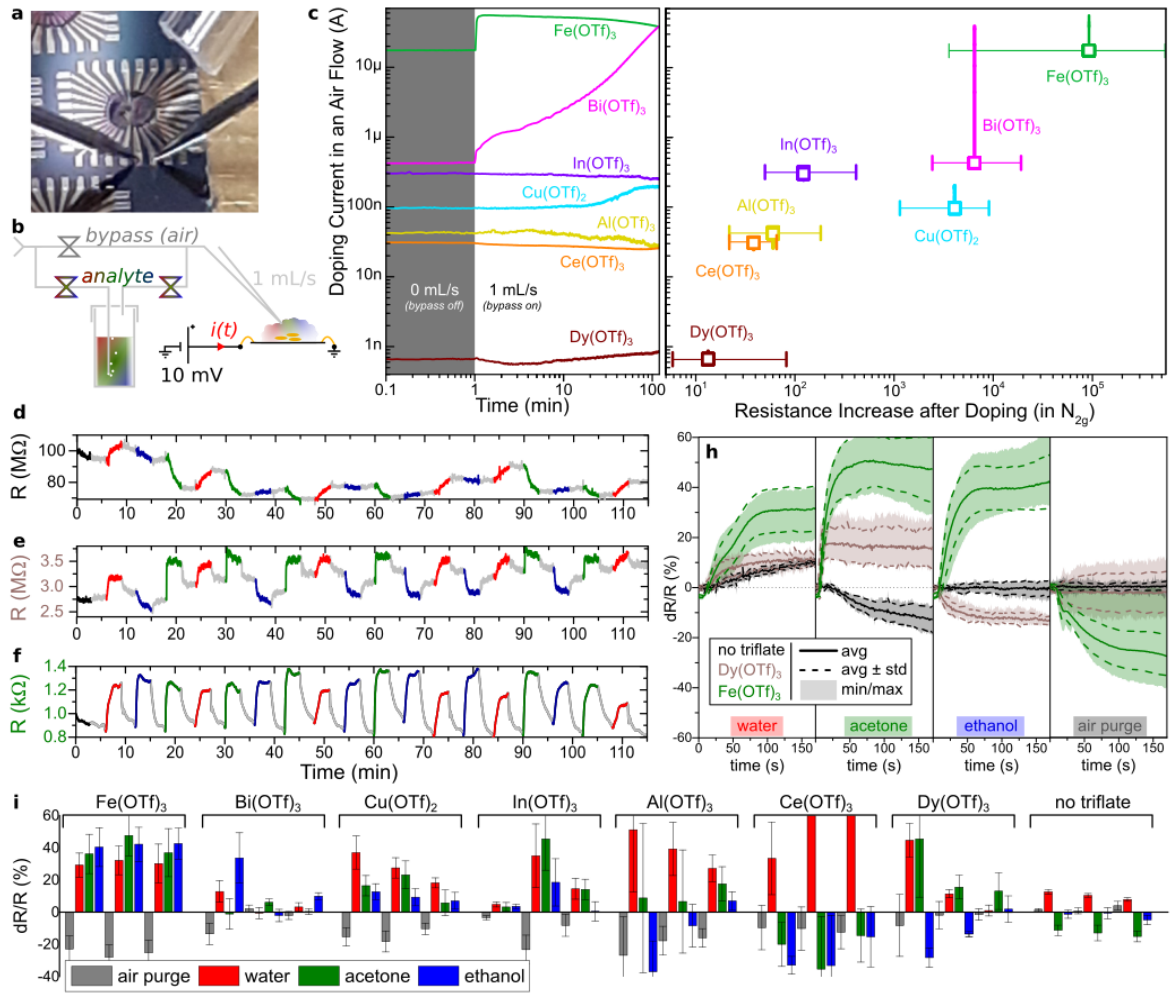

**Figure 2 | Molecular sensitivity assessment of the triflated materials under an air flow. a,** Photograph of the experimental setup. **b,** Schematic of the steady-aerodynamic (1 mL/s) employed to expose the triflated material devices under various conditions, under electrical measurment at 10 mV steady-voltage bias. **c,** Evolution over of the doping currents evaluated for each device before and after turing on the air flow, and systematic comparison of these dynamic currents with the rate of resistance increase of the materials evaluated under inert gas (the error bar represents the device property distribution of Fig. 1h). **d-f,** Illustrative example of three specific behaviours of the materials' sensitivity for untriflated (**d**), Dy(OTf)$_3$-sensitized (**e**) and Fe(OTf)$_3$-sensitized (**f**) P3HT upon iterative exposures of air saturated with water (red), ethanol (blue) and acetone (green). Each 3-min gas exposure is initiated by a 3-min air purge (grey), and the full sequences were recorded three minutes (black) prior before turning the flow. **h,** Statistic comparison of the different dynamics the four molecular environments electrically imprint on the three device materials. **i,** Sensitivity signatures of the molecular environments imprint on the eight different materials considering the relative resistance modulations recorded during the last ten seconds per exposure (three devices were evaluated for each material).

**Multivariate data analysis on combination of doped P3HT materials.**

On the nature of the doping/dedoping mechanism, the comparison of the different imprinted patterns on doped P3HT conductance shows that the mechanism is quite multi-parametric, not strictly ruled by one single physical property of the gas. To identify the nature of physical features that promotes different chemo-specificity for P3HT's molecular recognition, multivariate data analysis has been performed by mean of PCA. The analysis has been performed on a dataset of dynamical relative resistance modulation dR/R recorded for the eight different materials, three devices per material, with the output currents recorded overtime upon controlled exposure of the three different gases, exposed six sequences of 3 min per gas, and sampled at 1 Hz. To filter the amount of data to analyze, we first performed PCA on three different subsets of data that contains output currents (1) at the first 10 s of exposure named "transient state", (2) at the last 10 s of exposure named "steady state", and (3) an identical sampling over the full range named "whole regime" (Figure 3a - inset). In each analysis, we observe that the evolution of the cumulative variance in the sub-datasets is quite comparable for all three cases, with 60-70% of the total variance contained in the first two principal components (Figure 3a). Comparing the (PC1 ; PC2) projections for all three cases, we chose to use the "steady state" dataset which shows the best separation of data by nature of the gas (Figure 3b, the two other projections for "whole regime" and "transient state" are available as a supplementary information). The first two principal-components plane shows that data projections can be linearly classified by exposed gas (purple dash line in Figure 3b). The hierarchical clustering heatmap (Figure 3d) shows that each data are first clustered by gas nature and then by exposure sequences, which indicates that each of the six clusters in each of the PCA plot ellipse represents data of different 3-min sequences. It also indicates that Bi(OTf)$_3$-doped and Fe(OTf)$_3$-doped P3HT devices have a specific contribution for discriminating gases compared to the other dopants. From the PCA loadings analysis, one can observe that the contribution of these two dopants on PC1 is positive while other materials' loading are negative and comparable (Figure 3c): One can extrapolate from PC1 that Fe(OTf)$_3$ and Bi(OTf)$_3$ promotes discriminating water (PC1 < -2.5 in Figure 2b) from acetone and ethanol (PC > -1 in Figure 2b). This property can be related with the observation made previously on the hygroscopy of these strongest electron acceptors that turns to be an advantage to recognize water from volatile organic compounds. Similarly, PC2 promotes discriminating acetone (PC2 < -2.5 in Figure 2b) from water and ethanol (PC2 > -0.5 in Figure 2b). Pristine P3HT or doped with Ce(OTf)$_3$ are the only materials that have negative loadings on PC2 for all their three devices. This might be explained by the fact only those two materials showed substantial over-doping by acetone (negative resistance modulation), while all the other materials tend to reduce doping by the presence of this gas (Figure 2i). The squared loadings heatmap in Figure 3e shows that all eight materials have non-negligible contributions in the first two principal components, implying that all of them are relevant to discriminate gases with PC1 and PC2. However, one can see that Fe(OTf)$_3$ and Bi(OTf)$_3$ positive contributions in PC1 are not as significant as the other dopants' (Figure 3e), suggesting that their particular hygroscopy is not too significant in PC1's variance. A reason for this may come from the relative lack of selectivity/specificity that was observed in the case of those two materials: Fe(OTf)$_3$ shows large resistance modulations which are almost identical for all three gases, and Bi(OTf)$_3$ has poor reproducible sensitivity on the three devices (Figure 2i). The squared PC loadings for pristine P3HT are rather distinctive from the triflated materials ones, as most of it density is localized on PC1, PC2 and the least five principal components (Figure 3e). At the opposite of the other materials which show substantial densities on all principal components, it suggests the sensitivity model of pristine P3HT is much simpler than triflated systems', with only two main sensitivity parameter features.

We investigated on the minimum amount of materials required to successfully discriminate the three gases to simplify a sensing array and co-integrate the lowest number of necessary materials for an efficient classification (Figure 3g-h). To do this, the database dimensions composed of 24 devices responses was reduced to n (1≤n≤8), where the n devices are all made of different P3HT materials.

To evaluate the recognition performances of the lower-dimensional sensing arrays, we defined a two dimensional output vector as:

$$OUT_k = \sum_{i=1}^{n} \overline{x_i^{PCk}} \left[\frac{dR}{R}\right]_i \quad ; \quad k = 1 \text{ or } 2$$

Where $\overline{x_i^{PCk}}$ is the average value of the PC$_k$ loadings $x_i^{PCk}$ displayed in Figure 3c for the three devices of i material, and $\left[\frac{dR}{R}\right]_i$ the relative resistance modulation recorded for the selected device response i.

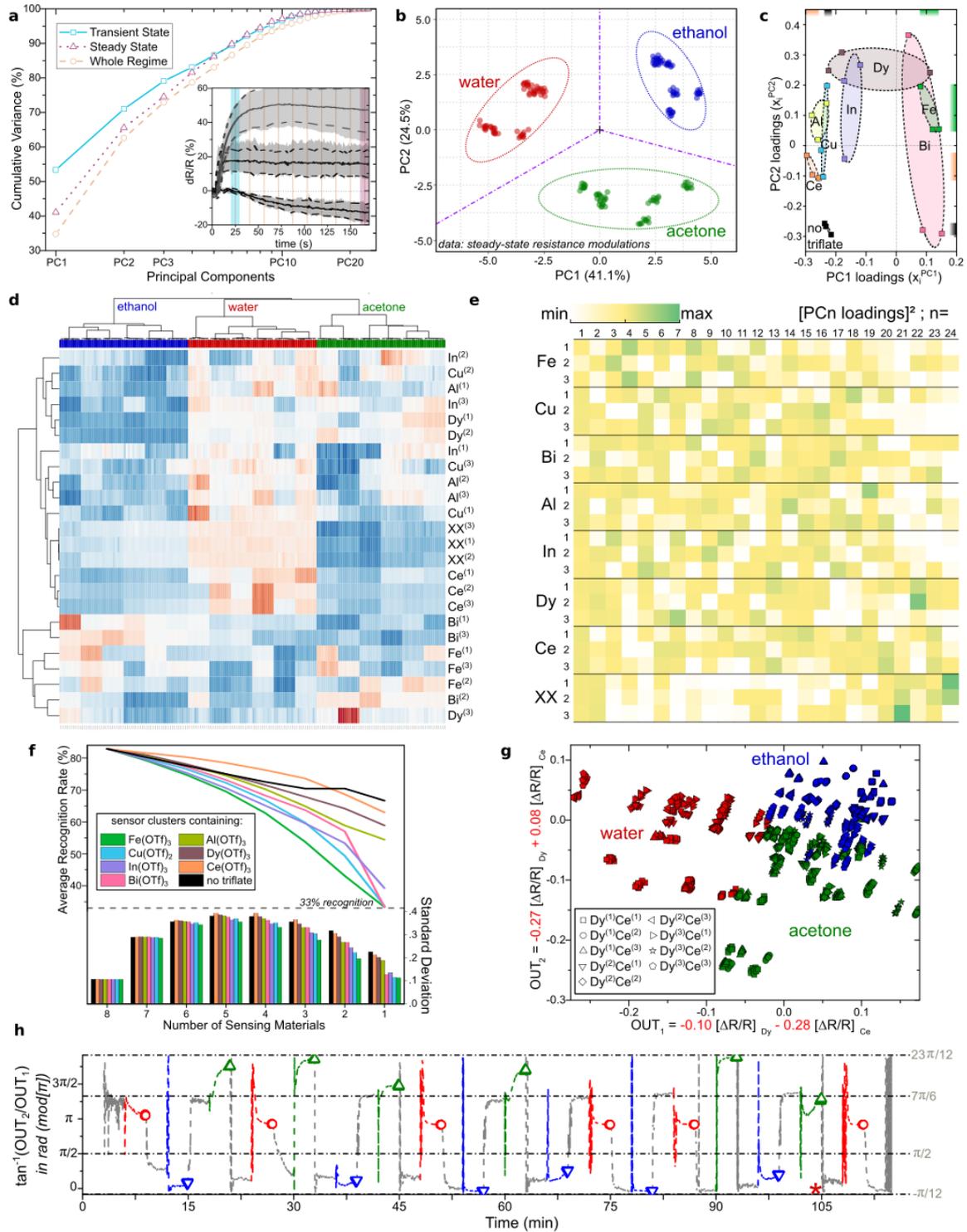

**Figure 3 | Multivariate Data Analysis & Chemo-recognition. a,** Cumulative Variance for the Principal Component Analysis (PCA) considering different datasets of 10 points per gas sequence recorded at different times after gas exposure: between 20 and 29 s each 1 s (labelled "transient state"), between 170 and 179 s each 1 s (labelled "steady state") and between 25 and 178 s each 18 s (labelled "whole regime") - the inset illustrates the position of the different data used for the three analysis). **b,** First two principal components graph for the PCA performed with the "steady state" dataset. **c,** PCA loadings for the first two principal components showin the different material contributions. **d,** PCA Heatmap with the "steady-state" dataset. **e,** Relative contributions of the different materials on the square loading of the PCA performed with the "steady-state" dataset. **f,** Recognition performance for different sizes of sensing materials/device clusters upon ternary linear classification ternary classification. **g,** Two-dimensional sensing response for sets of two sensors composed only of $Dy(OTf)_3$ and $Ce(OTf)_3$. **h,** Illustrative example for using $\tan^{-1}(OUT_2/OUT_1)$ as 1-dimensional biparametric information feature to recognize the three gases with only two $Dy(OTf)_3$ and $Ce(OTf)_3$ devices (here in the case of devices $Dy^{(2)}$ and $Ce^{(2)}$ that show perfect classification at the exception of one gas sequence marked with a *).

As calibration rules to classify the outputs, three linear conditions are set on the polar angle ($OUT_1$;$OUT_2$) coordinates, based on the purple dashed lines represented in the PCA plot in Figure 3b:

$$\forall\ OUT_1 > 0, \quad if\ \tan^{-1}\left(\frac{OUT_2}{OUT_1}\right) > -\frac{\pi}{12}, \ then\ class = \text{"ethanol"}$$

$$else\ \forall\ OUT_1 < 0, \quad if\ \tan^{-1}\left(\frac{OUT_2}{OUT_1}\right) < +\frac{\pi}{6}, \ then\ class = \text{"water"}$$

$$else\ class = \text{"acetone"}$$

Based on the comparison of the class value suggested by the aforementioned calibration rules, and the actual identity of the gas that is exposed on the randomly picked devices, an average recognition rate is evaluated for all the different combinations on n materials (Figure 3f). We evidenced that some of the $4^8$ = 65536 combinations of devices shows better recognition rates than others, for which an average value of recognition rate has been determined by number of sensing materials (n), and by kind of triflate deposited (Figure 3f). It was observed in all cases that the more materials are included in the analysis, and the better is the recognition: From 82% on the analysis of n = 8 device sensing array containing all P3HT materials, down to 33% in the case only one $Fe(OTf)_3$ or one $Bi(OTf)_3$ device is used. In the specific case of $Fe(OTf)_3$ that shows that de-doping is almost identical in each case for water, acetone and ethanol (Figure 2f,i), the lowest recognition rate of 33% shows that using only the most performing dopants does not ensure the algorithm to be able recognizing better than stochastically. Although pristine P3HT's resistance was the highest and its modulations toward gas exposure were much lower than for $Fe(OTf)_3$ (Figure 2d,f,i), using only one untriflated device allows recognizing gases by 67%, using its versatility to be either over-doped by water, or de-doped by acetone (Figure 2d,f). More interestingly, the recognition performances for n = 1 single device array shows that the algorithm recognize better with mild-doped systems than with strong acceptors (Figure 3f). Increasing the size of the array n increase furthermore the recognition performances, specifically to the combinations of materials included in the analysis: With n = 2 devices, any of the nine combinations of $Ce(OTf)_3$ with $Dy(OTf)_3$ devices guaranties at least 73.3% recognition rate. In case of n = 3, the two best sets identified are $Ce(OTf)_3$ with $Dy(OTf)_3$ with $Bi(OTf)_3$, or $Ce(OTf)_3$ with $Dy(OTf)_3$ with untriflated-P3HT, with a guaranteed recognition performance of 72.2%. As $Ce(OTf)_3$ with $Dy(OTf)_3$ is the best set of n = 2 materials to recognize the three gases, the output for the combinations of device pairs is displayed in Figure 3g. The data remains linearly separable according to the calibration rules defined from the original PCA plots (Figure 3b). Out of the nine combinations, Figure 3h displays the behavior of the output over time for one specific combination of $Ce(OTf)_3$ and $Dy(OTf)_3$ devices that fails recognition of acetone on only one sequence (marked by a red star). Surprisingly, the graph shows that although the PCA was performed with only the last ten seconds of data of each sequences (marked with scatters on the line), the calibration rule seems to be adequate to recognize also the data all along the whole transient, which were not included in the calibration dataset for the PCA.

**Dopant's coordination chemistry in conducting-polymer gas sensing.**

Multivariate data analysis shows that sensitizing two cells coated of a same polymer but doped by different lanthanide triflate promotes different resistance modulations towards exposures of water, ethanol and acetone vapors, substantially enough so an array of them can help to differenciate three different environments. This confirms the participation of the whole triflate salt in the P3HT doping mechanism. It also tells that cations sensitize P3HT's by interfering with the dopant/semiconductor charge-transfer (CT), as the transduction mechanism triggering the resistance change of the device. When comparing the potentials of cations redox couples in water to the effect of their corresponding triflates on P3HT's conductance in inert atmosphere or in air, correlations between cations redox chemistry and triflate's doping can be noticed (Figure 4a). As the oxidation potential is a direct indicator of an electron acceptors' strength,[**Djurovich2009**] the correspondence of the three series in Figure 4a suggests that P3HT's CT with triflate salts is yielded by the depth of the triflates' lowest unoccupied molecular orbital (LUMO). However, correlations with redox activity of free ions in water do not presume that dry triflates undergo single-electron CT with a P3HT matrix, nor with any electron-donating specie present in its environment. As a clue, one can evidence that the molecular sensitivity of the triflated P3HT devices has followed very distinctive resistance modulation signatures upon exposure of electron donating vapors that does not explicitly respect the three aforementioned series (Figure 4b). For instance, we noticed that although $Fe(OTf)_3$ and $Bi(OTf)_3$ are particularly strong acceptors, hygroscopic materials and strong dopants for P3HT, $Fe(OTf)_3$ is far more sensitive to gases than $Bi(OTf)_3$ (Figure 2i, 4b).

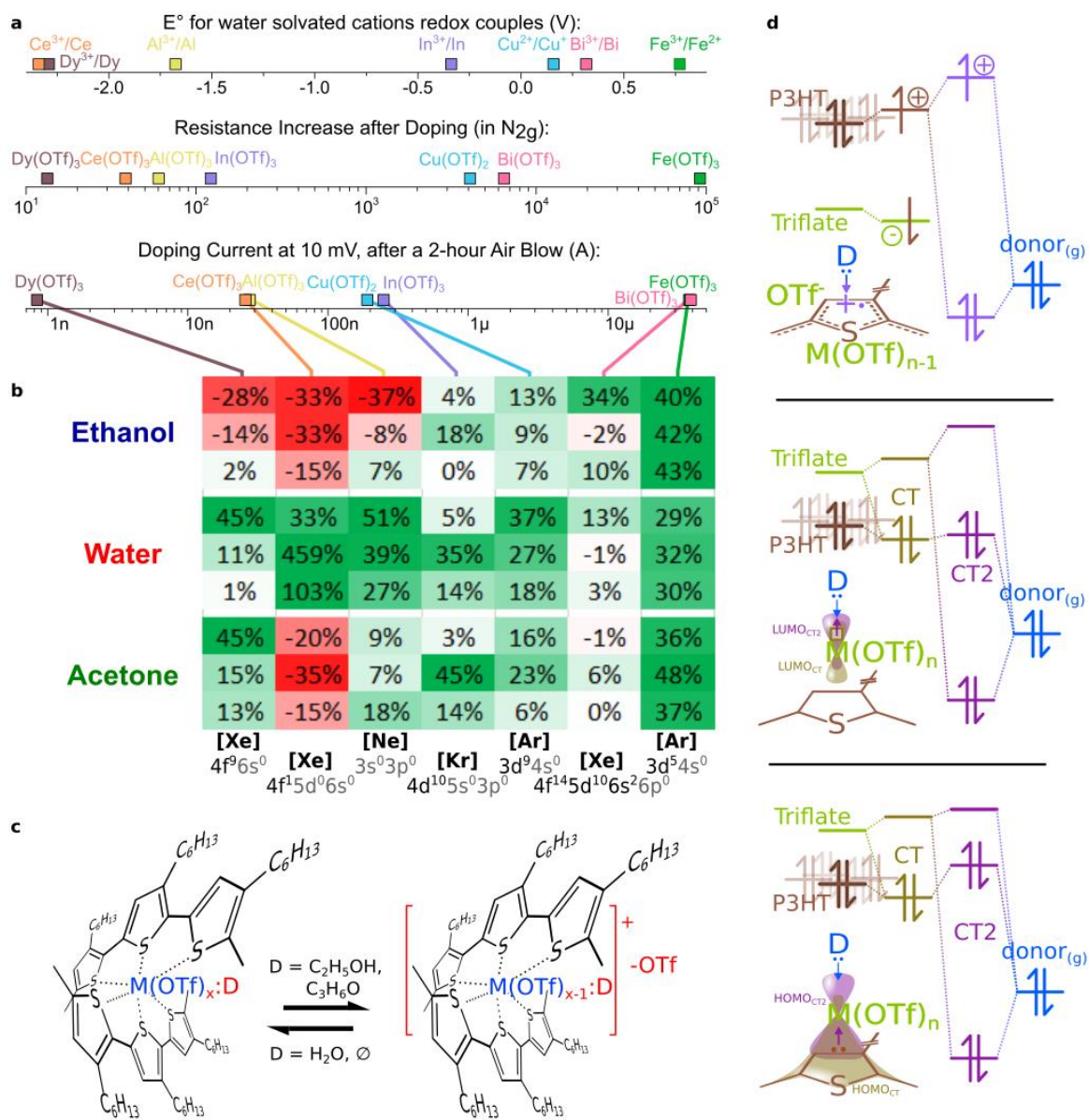

**Figure 4 | Chemospecific Detection/Transduction Mechanism(s) for Molecular Recognition. a,** Comparison of different electronic values obtained for the different triflated P3HT material devices recorded both under nitrogen and after operation in air, to the redox properties of the metal center as a free oxidant solvated in water (values from [**Dean1999**]). **b,** Relative resistance modulation (average values) recorded upon exposure of ethanol, water and acetone for each of the three devices for all the seven triflated P3HT materials (under, the electronic structure of the atomic orbitals for each triflate cationic metal/main-group-element core). **c,** A possible mechanism derived from $Ce(OTf)_3terpy_2$ adducts with volatile organic compounds to illustrate the particularly pronounced dedoping/overdoping behaviour for $Ce(OTf)_3$ on P3HT with donor gases D. **d,** Overall mechnisms undergoing by the doping/dedoping of the triflated-semiconductors on their frontier orbitals based on the strength of the charge transfert between the triflate salt and P3HT on the one hand, and with the lone electron pair donating gases on the other hand.

More interestingly, we noticed that weaker dopants can exhibit versatile resistance modulations according to the element of the salt's cation (Figure 4b): to be either systematically de-doped by any of the three gases (Cu(OTf)$_2$ and In(OTf)$_3$), or to show ethanol-specific over-doping (Dy(OTf)$_3$ and Al(OTf)$_3$) or by both ethanol and acetone (Ce(OTf)$_3$). The fact that some gases can over-dope some triflated-P3HT is a clear indicator that gases can selectively add-up to the electron withdrawing effect of the dopant and promote P3HT doping furthermore. It suggests that gases can participate to the dopant/P3HT CT by coordination with the cation, forming tripartite CT complexes gas/dopant/semiconductor. Such hypothesis opens different pathways for doping modulation under the hybrid CT complex framework: [**Salzmann2012**][**Salzmann2016**][**Méndez2013**][**Méndez2015**]

As first case scenario (upper scheme in Figure 4d) for strong redox-active electron-acceptors (like Fe(OTf)$_3$), the transfer of charge is integral because of M$^{\alpha}$ that can reduce to M$^{\alpha-1}$ by releasing one OTf$^-$ through one electron CT to oxidize polythiophene.[**Gueye2016**] In this case, the monomer units can bear an integral charge. Coulombic interaction between the charge carriers and triflate anions can only be further stabilized by the presence of polar gases, and therefore, this mechanism can only lead to systematic de-doping (like evidenced for Fe(OTf)$_3$ in Figure 2i). As this is specific to strong redox active dopants, one can imagine comparable behavior for being systematically de-doped by gas with other performant p-dopants. In the case of Bi(OTf)$_3$, we observe that despite its strong doping effect, the salt cannot freely oxidize three neighboring thiophene units through single-electron CT to undergo from Bi$^{3+}$ to Bi$^0$ (up to our knowledge, no chemical reduction from Bi(III) to Bi(II) has ever been reported for mononuclear bismuth molecular compounds).[**Coughlin2021**][**Dikarev2004**]

As a second case (middle scheme in Figure 4d), hybrid dopant/semiconductor CT can form a tripartite CT2 with a lone-pair electron-donating gas for which interactions with the dopant can reversibly kill partly its doping effect on the semiconductor. Such electron transfer can be promoted by a gas with the highest-occupied molecular orbital (HOMO as an oxygen doublet for water and ethanol or a C=O double bond for acetone) that strongly overlaps with the lowest unoccupied molecular orbital of the dopant/semiconductor CT on the electrophilic site of the triflate salt's core. As strong de-doping interferers, nucleophilic gases can overlap most of LUMO's CT on the triflate to diminish its doping effect on P3HT by an increased bandgap CT2 compared to the original P3HT/triflate CT. In our experiments, this case is specific of the effect of water on mild dopants which specifically de-dope triflated P3HT by coordinating the metal/main-group-element core and strongly overlapping the CT's LUMO as sterically unhindered gas.

In a third case (lower scheme in Figure 4d), the tripartite CT2 can show enhanced doping effect on P3HT in case of stronger electron withdrawing effect on the triflate to enhance the CT's HOMO depopulation from the thiophene site. Evidenced in the case of ligand substitution of arylamine doping by coordination compounds, strongly acidic ligands can promotes better doping on a Lewis acid than weaker ones.[**Schmid2014, Pecqueur2016**] Here in case of weak triflates (such as lanthanide-metal salts Ce(OTf)$_3$ and Dy(OTf)$_3$) compared to stronger but hybrid dopants (such as transition-metal salts : Cu(OTf)$_2$ or In(OTf)$_3$), the effect of molecular gases on the dopant can reverse, from inhibiting to promoting, depending on the original CT rate of the metal triflate with P3HT and also on the capability of the gas to access the nucleophilic sites of the dopant. A specific point should be highlighted on the fact that ethanol can over-dope the three weakest dopants while acetone over-dopes only Ce(OTf)$_3$.

The case of Ce(OTf)$_3$ is particularly unique as its sensitivity to water was particularly strong to de-dope P3HT (33% to 459% of resistance increase) while it is the only one that shows over-doping by both ethanol and acetone (15% to 35% output current increase). As evidenced by X-ray crystallography in the case for lanthanide triflate terpyridine adducts (M(OTf)$_3$:terpy$_2$), the adjunction of lone-pair donating solvents (L = acetonitrile or pyridine) can promote the oxidation of a cationic adducts by the release of a triflate anion upon chelation of L on M as [M(OTf)$_2$:terpy$_2$]$^+$OTf$^-$ with M=Ce,[**Berthet2005**] further observed for other lanthanide triflates M(OTf)$_3$ and for other tridentate aromatic systems[**Escande2009**] If oligothiophene trimeric units have an analog role on the formation of such bi-stable adducts by the presence of lone-pair donating molecule in the gas phase, such mechanism can justify the particular sensitivity of Ce(OTf)$_3$ to ethanol and acetone compared to the other triflate salts (Figure 4c). Here should be noted that although the terpy/triflate adduct stabilizes through single electron CT and releases OTf- upon addition of volatile solvents, redox potential of Ce$^{3+}$/Ce in water is far too low (Figure 4a) to suppose that Ce(OTf)$_3$ itself is strong-enough electron-accepting to undergo Figure 4c's upper mechanism. However, one can suppose from the third case (Figure 4c lower scheme) that Ce(OTf)$_3$:P3HT HOMO's CT pull-up can promote reducing its bandgap low enough so it becomes comparable to [Ce(OTf)$_2$:P3HT]$^+$OTf$^-$ Coulomb energy, where the chelation of specific electron donating gases can shift the equilibrium between hybrid CT to integer CT via single-electron CT.

All over, it should be mentioned that all proposed schemes or combinations of them do not take into account differences between closed- and open-shell dopants (for which electronic spin induces other orbital shifts from the Ligand-Field Theory) nor the spdf nature of the metal/main-group-element for which their electrophilicity differs and unfilled frontier-orbitals hybridize specifically with P3HT's HOMO on the one hand and with gases' on the other hand. Figure 4b specifies that all salts cation elements belong to different blocks of the periodic classification. Up to now, no clear property has been evidenced, except that both f-block dopants are mild but chemo-specific to molecular gases, while the three d-block dopants are better dopants but show only de-doping. No clear correlation between both p-block element dopants has been evidenced, or between the three closed-shell systems and the four open-ones. This opens a large new field of investigations to further understand the correlations between chemo-sensitivity of doped systems and molecular properties of dopant/semiconductor complexes, leading to the conception of broader chemo-perception fields for various sensing platforms that aim environmental sensing through molecular recognition.

## Conclusion

A new class of mechanisms has been identified to sensitize semiconducting materials in a way they become chemo-specific to molecular gases, with conductimetric sensing micro-arrays for multivariate chemo-recognition. Using different triflate salts on poly-3-hexylthiophene leads to chemo-specific doping, where the molecular identity of an exposed gas allows imprinting chemo-specific patterns of polymer conductance modulations. Thanks to the interaction of the gas with the triflate acceptor that dopes the polymer, a same polymer is able to discriminate water from acetone from ethanol in ambient conditions with a unique conducting polymer, but doped with at least two triflates. More than simplifying electronic noses material co-integration for disease biomarker detection or air quality monitoring, the experimental results coupled with multivariate data analysis highlights the involvement of molecularly-specific tripartite mechanisms between organic semiconductor, a Lewis-acidic p-dopant and an electron-donating environment. With future developments to be made on dopant/semiconductor co-integration, the demonstration of molecularly-specific tripartite charge-transfers in organic electronic materials is a unique technological tool to integrate electronic nose on portable IoT devices, but also an enthusiastic opportunity to better understand the role of molecular orbitals in the organic semiconductor charge-transfer doping.

# Methods

**Device Fabrication:** Devices were fabricated by electron-beam lithography to structure the electrode and passivate their contact lines (based on a process reported elsewhere)[**Pecqueur2018b**], while the doped polymer was wet-deposited successive drop casting: The electrodes active area were patterned by lift-off from a 50-nm thick thermally-evaporated gold layer and 10 nm titanium to adhere on a 200 nm $SiO_2$-passivated Si wafer die. The channel length defined as the distance between both interlaced electrodes was 400, 800, 1200, 1600 and 2000 nm all over the 50-nm thick gold shape. The 300-nm contact lines were structured similarly via lift-off, and covered by a 2 µm-thick Parylene C passivation layer. A 28-µm diameter round lithographically-patterned cavity was structured all-the-way through the passivation layer by a $O_2$-plasma etching. Regio-regular P3HT and dopants were purchased and used without further purification. First, 10 mg/mL of pristine poly(3-hexylthiophene) solutions were formulated in a (7:1) chlorobenzene/ethanol blend. The deposited material was dried at 70°C during 2min in air before evaluating its electrical performances as "before doping", prior depositing the triflate salt on top. Second, 10 mg/mL concentrated triflate salt solutions were formulated in pure ethanol and deposited on top of dry P3HT similarly. Another 2min annealing at 70°C in air was done to help increasing the material's conductivity "after doping" and dry the ethanol.

**Electrical Characterization:** Electrical characterizations were performed on an Agilent 4155 parameter analyzer in $N_2$ glovebox and in air. The exposure of the different gases was controlled manually by three valves as depicted previously. Due to the experimental setup, up to five seconds of cycle asynchronization can be considered, between the gas label stamped on the recorded data and the actual composition of gas flowing over the device at a given time.

**Data Analysis:** Principal component analysis has been performed with both the online open access tool Clustvis [**Metsalu2015**] and the Python library Scikit-Learn [**Pedregosa2011**] (both techniques gave exactly the same results). Input data for the PCA were the relative resistance modulation calculated from the method described with the output current data, as is, without filtering nor smoothing. PCA data scaling by unit variance and computed by singular value decomposition.

# Supporting Information

Supplementary information is available in the online version of the paper.

# Acknowledgements

The authors thank the French National Nanofabrication Network RENATECH for financial support of the IEMN cleanroom. We thank also the IEMN cleanroom staff of their advices and support.

# Competing Interests

The authors declare no competing interests.

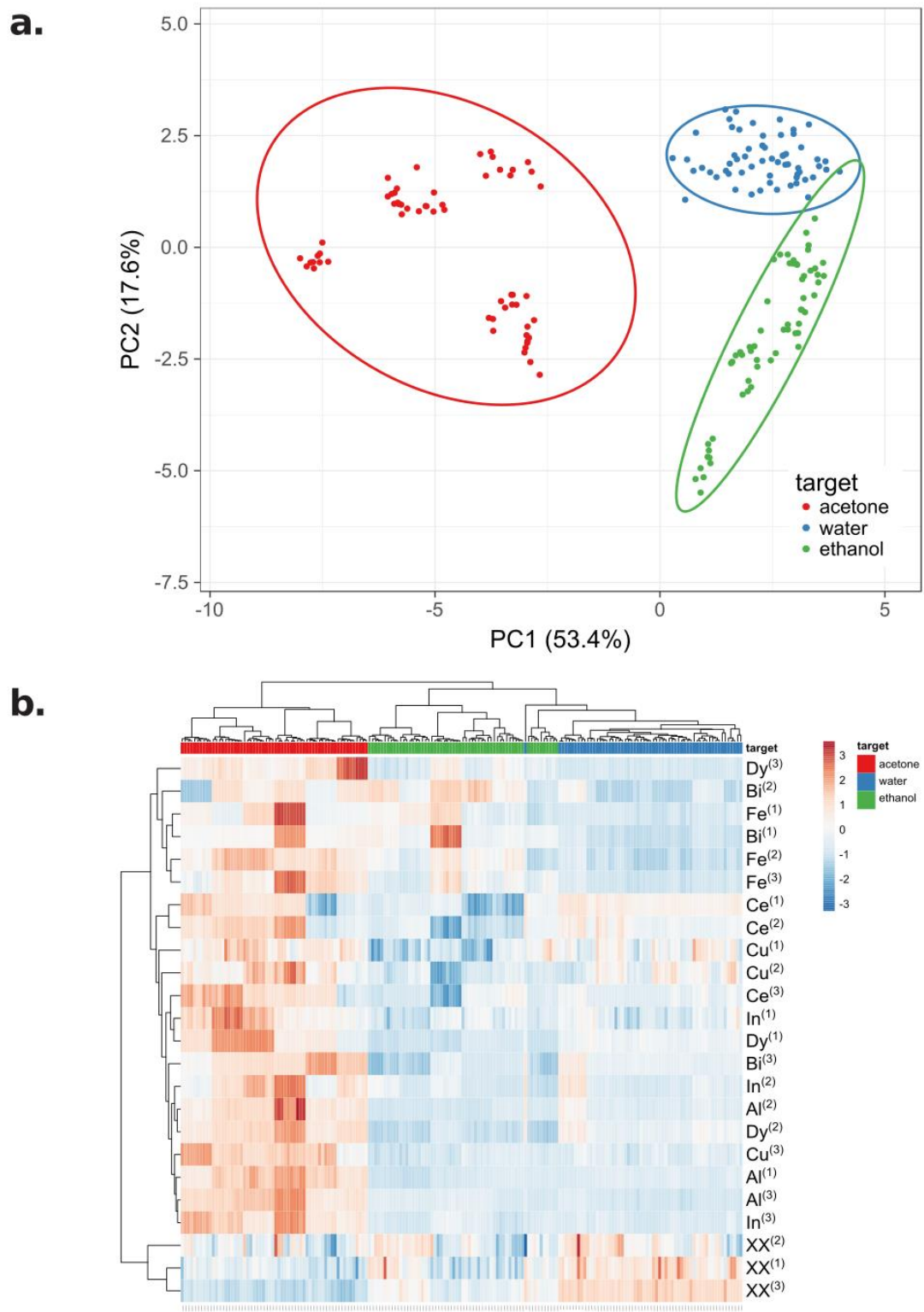

**Figure S1 | Principal Component Analysis on ten "transient-state" sensor data per sequences of gas exposure (from t = 20 to 29 seconds every 1 seconds after gas exposure) a,** 2-dimensional projection of the 180 points (3 gas, 6 sequences, 10 samples) of 24-dimensional data (8 materials, 3 devices each) on the first two principal components plane. **b,** Hierarchical heat map cluster showing data inter-relashionship between the nature of the gas (three targets) and its impact on the 24 different devices of 8 different materials.

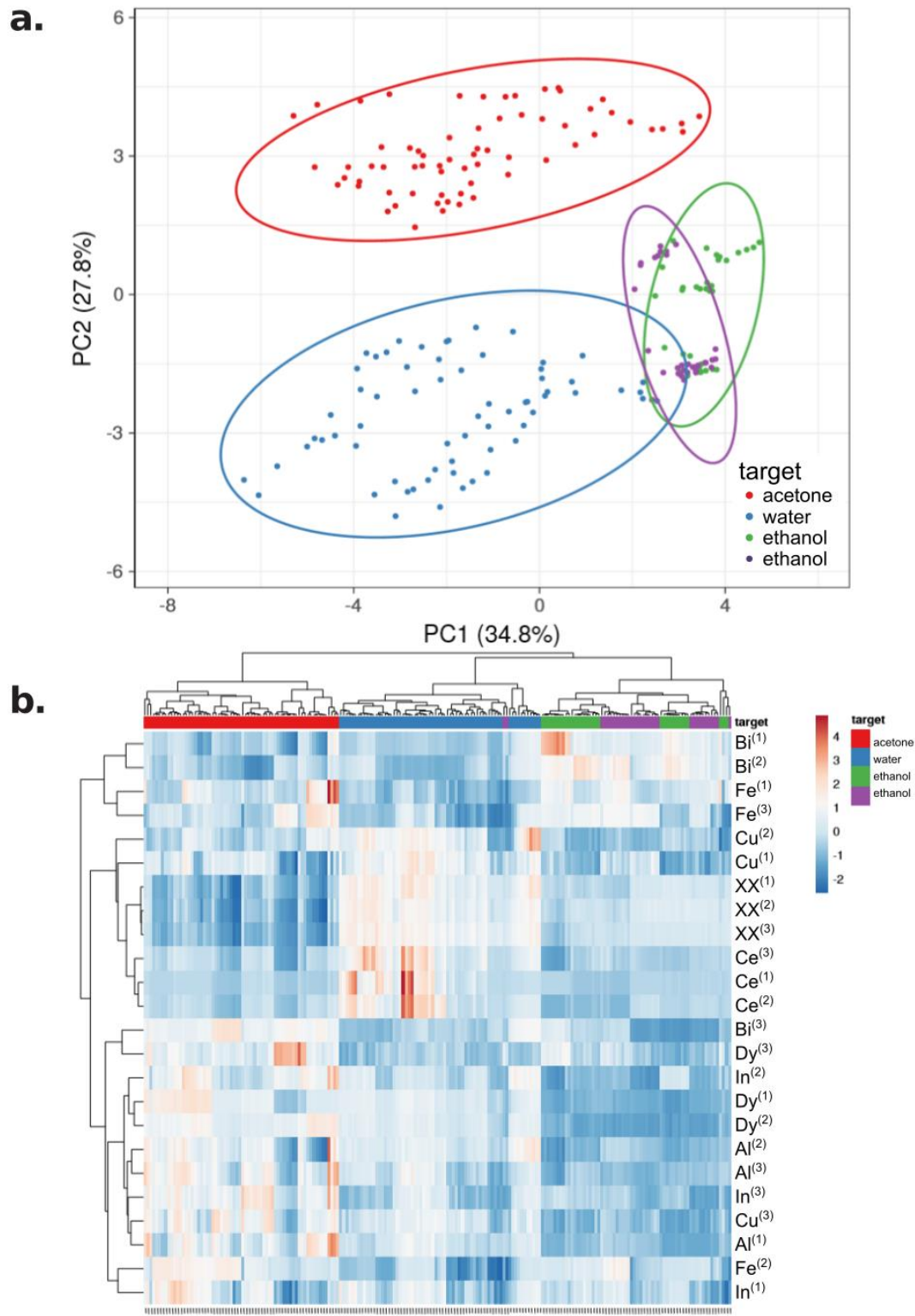

**Figure S2 | Principal Component Analysis on ten "whole regime" sensor data per sequences of gas exposure (from t = 25 to 178 seconds every 18 seconds after gas exposure) a,** 2-dimensional projection of the 180 points (3 gas, 6 sequences, 10 samples) of 24-dimensional data (8 materials, 3 devices each) on the first two principal components plane. **b,** Hierarchical heat map cluster showing data inter-relashionship between the nature of the gas (three targets) and its impact on the 24 different devices of 8 different materials.